\documentclass[11pt,notitlepage]{article}
\usepackage{graphicx}

\begin{document}

\title{Further Comments on a Vanishing Singlet Axial Vector Charge }
\date{ IFUP-56/98, UMSL-98-9, FTUV/98-101 and IFIC/98-102}
\author{T. P. Cheng$^{\ast }$, N. I. Kochelev$^{\dagger }$, and V. Vento$%
^{\diamond }$ \\
$^{\ast }${\small Dept. of Physics and Astronomy, Univ. of Missouri,
St.Louis, MO 63121, USA}\\
$^{\dagger }${\small 
Dipartimento di Fisica, Universit\`a di Pisa , I-56100 Pisa, Italy}\\ 
{\small and Joint Institute for Nuclear Research, Dubna, Moscow
Region, 141980 Russia}\footnote{Permanent address}\\
$^{\diamond }${\small Dept. de F\'{i}sica Te\`{o}rica, } {\small Univ. de
Val\`{e}ncia, E-46100 Burjassot, Spain}}

\maketitle

\begin{abstract}
The recent suggestion of a vanishing flavor-singlet axial charge of nucleon 
due to a
nontrivial vacuum structure is further amplified. A perturbative QCD
discussion, applicable for the heavy quark contributions, relates it to the
physics of the decoupling theorem. It is also shown that $g_{A}^{0}\simeq 0$
leads to a negative $\eta ^{\prime }$-meson-quark coupling, which has been
found to be compatible with the chiral quark model phenomenology.
\end{abstract}

\section{Introduction}

There are two popular descriptions of a suppressed quark contribution to the
nucleon  spin. One approach suggests that it is due to the (low energy)
non-perturbative QCD  depolarization  of quarks inside 
nucleon\footnote{%
For a recent review of the non-perturbative spin/flavor study in general,
and chiral quark model in particular, see \cite{CL-spin98}.}. Another
approach attributes this suppression to a possible gluon polarization,
contributing via the axial anomaly, which reflects the high energy
regularization of certain quark loop diagram\cite{glue-spin}.

Understanding the nucleon spin problem is related to understanding the its 
flavor 
singlet axial charge ,  $g_{A}^{0}$. The latter for a spin 
$\frac12$%
\thinspace target is defined by the formula 
\begin{equation}
<p^{\prime }|J_{\mu 5}^{0}|p>=g_{A}^{0}(Q^{2})\bar{P}\gamma _{\mu }\gamma
_{5}P+g_{P}^{0}(Q^{2})q_{\mu }\bar{P}\gamma _{5}P  \label{mat}
\end{equation}
where singlet axial vector current is not conserved due to the anomaly: 
\begin{equation}
\partial _{\mu }J_{\mu 5}^{0}(x)=2i\sum_{i}m_{i}\bar{q}\gamma _{5}q+2N_{f}%
\frac{\alpha _{s}}{8\pi }G_{\mu \nu }^{a}\widetilde{G_{\mu \nu }^{a}}.
\label{anom}
\end{equation}

In a recent paper by two of us (NIK and VV)\cite{kochvento} it was argued
that the properties of the QCD vacuum lead to a vanishing of 
the flavor-singlet axial vector charge:

\begin{equation}
g_{A}^{0}\simeq 0.  \label{vanishing-gA}
\end{equation}
We also discussed the relation of our result with the above descriptions.

Let us
recall the non-perturbative mechanism behind  
the suppression of $g_{A}^{0}$. Once the anomaly is 
interpretated as the motion of the Dirac levels of definite chirality
\cite{Shifman}, the suppression comes about due to the cancellation between the
infrared (IR) and ultraviolet (UV) contributions, which arise in order to 
preserve the 
axial-charge of the vacuum. Thus
the vanishing of the axial-charge  holds in
the flavor-singlet channel, independent  
of the target: be it hadron or constituent quark. The possibility of a
target-independent suppression of the singlet nucleon axial charge has also
been discussed previously using different approaches\cite{BEK,eta-V,eta-EGK}.

We have also shown how this phenomenon is realized in the instanton model of the
vacuum. An instanton dominated non-perturbative QCD is a plausible mechanism 
for the low energy depolarization as the 't Hooft determinantal interaction
\cite {tHooft} flips the quark helicities\cite{Koch}. This interaction was
obtained by taking into account the contribution of only the quark 
zero modes in the instanton field.

The axial anomaly is related to the UV
regularization in the perturbation theory and therefore to the non-zero quark
modes.  Once the zero mode contribution and the non-zero mode contribution
to the singlet axial charge are considered, the suppression occurs 
\cite{kochvento}. This is how
the instanton vacuum realizes our general result.

In this note we shall provide more arguments which support  the validity  
of the singlet axial-charge suppression due to the properties of 
the QCD vacuum. First, in Sec.2 we present a simple version of
this cancellation in the case of heavy quark contribution to the axial
vector charge, showing how this cancellation comes about in the
perturbative approach. This is in accord with the expectation that heavy
particles must decouple from low energy physics\cite{AC-decoupling}. In
Sec.3 we then show that the vanishing of the flavor singlet axial charge
leads to the negative value for the $\eta ^{\prime }$-quark coupling, which
has been found in an earlier publication\cite{CL95} to be compatible with
chiral quark model phenomenology.

\section{The perturbative version for heavy quarks}

We nest present a perturbative version of the result in Eq.(\ref
{vanishing-gA}): the decoupling of the heavy quark contribution to the
target spin. In accordance with a decoupling theorem one would  expect that 
any heavy quark $Q$ makes a vanishingly
small contribution to the axial vector charge since
the involved energy is much  smaller than the heavy quark rest energy $m_{Q}.
$

This can be demonstrated, for example\cite{CL-ricemtg}, by following the
heavy quark expansion idea of Witten\cite{Witten}. When we integrate out the
heavy quark field $Q$, because the presence of the quark mass $m_{Q}$, the
pseudoscalar density\ $im_{Q}\bar{Q}\gamma _{5}Q$\ \ in the current
divergence does not vanish, but gives rise to a dimension-four operator: 
\begin{equation}
im_{Q}\bar{Q}\gamma _{5}Q\longrightarrow -\frac{\alpha _{s}}{8\pi }G_{\mu
\nu }^{a}\widetilde{G_{\mu \nu }^{a}}+O\left( m_{Q}^{-2}\right) 
\label{heavy-expansion}
\end{equation}
The relevant diagram is just the pseudoscalar triangle diagram with two
gluon legs. This minus sign, which leads to the above-mentioned cancellation
by the anomaly term in Eq.(\ref{anom}), can be understood this way. The
anomaly term results from regulating the UV divergence. This can be
implemented by introducing into the theory a regulating fermion $R$ with a
large mass $M_{R}$. This breaks the chiral symmetry and gives rise to a mass
term in the divergence of the axial vector current\ \ $iM_{R}\bar{R}\gamma
_{5}R$. \ In the limit of $M_{R}\rightarrow \infty ,\;$this mass term is
then transformed (when integrating out the $R$ fermion) into the anomaly
term: 
\begin{equation}
iM_{R}\bar{R}\gamma _{5}R\longrightarrow \frac{\alpha _{s}}{8\pi }G_{\mu \nu
}^{a}\widetilde{G_{\mu \nu }^{a}}  \label{mass-to-anom}
\end{equation}
These two cases of the heavy quark and regulating fermion are exactly the
same, the only difference being that the regulating fermion has the opposite
metric (its loop integral has an extra minus sign). This accounts for the
opposite signs on the RHS in Eqs.(\ref{heavy-expansion}) and (\ref
{mass-to-anom}), and a vanishing contribution to the axial current
divergence.

Then by using  Eq.(\ref{mat}) and the absence of the massless pole in the
form factor $g_{P}^{0}$ we come to the result in Eq.(\ref{vanishing-gA})
which can be regarded as an analog of the Goldberger-Treiman relation for
the flavor-singlet axial vector channel.

For the light quarks, perturbative QCD is not applicable. The low energy
non-perturbative QCD physics has to be modeled, \emph{e.g.,} by the
instanton dominance. What is argued in Ref.\cite{kochvento} is that in this
case the cancellation survives in the flavor singlet channel, even for the
light quarks, but only for  the matrix elements of the quark
and gluon operators in the right-hand side of Eq.(\ref{anom}), and not for
 the operators in Eq.(\ref{heavy-expansion}).

\section{Singlet coupling in the chiral quark model}

Since the U(1) symmetry is broken (by instantons), there must be mixing
among the pseudoscalar states, 
\begin{eqnarray}
\left| \eta ^{\prime }\right\rangle  &=&\;\,cos\Theta \left| \eta
^{0}\right\rangle +sin\Theta \left| \eta ^{8}\right\rangle   \nonumber \\
\left| \eta \right\rangle  &=&-sin\Theta \left| \eta ^{0}\right\rangle
+cos\Theta \left| \eta ^{8}\right\rangle   \label{mix}
\end{eqnarray}
where $\Theta \approx -18^{\circ }$ is the $\eta -\eta ^{\prime }$ mixing
angle. In the pole approximation for the matrix element of the
right-hand side of Eq.(\ref{anom}) the vanishing of the singlet axial
coupling in (\ref{vanishing-gA}) becomes 
\begin{equation}
-\frac{g_{\eta qq}}{{m_{\eta }}^{2}}sin{\Theta }+\frac{g_{\eta ^{\prime }qq}%
}{{m_{\eta }^{\prime }}^{2}}cos{\Theta \simeq 0}  \label{GT}
\end{equation}
or, 
\begin{equation}
g_{\eta ^{\prime }qq}\simeq tan{\Theta }\frac{{m_{\eta }^{\prime }}^{2}}{{%
m_{\eta }}^{2}}g_{qq}.  \label{coupl}
\end{equation}
By substituting in (\ref{coupl}) the experimental values for $\Theta $, $%
m_{\eta }$ and $m_{\eta ^{\prime }}$ one can find 
\begin{equation}
g_{\eta ^{\prime }qq}\approx -g_{\eta qq}.  \label{result}
\end{equation}
The negative sign reflects the negative mixing angle, which is related to
the resolution, via the instanton physics, of the axial U(1) problem,\ ${%
m_{\eta ^{\prime }}}^{2}\gg {m_{\eta }}^{2}$. The related nucleon result for 
$g_{\eta ^{\prime }NN}$ has also been discussed by other authors\cite
{eta-V,eta-HK}. The specific situation of a negative singlet-meson coupling
in an instanton dominated QCD (only for the zero-mode determinantal
interaction) has been discussed in Ref.\cite{CL98-instanton}.

In an earlier publication\cite{CL95}, it has been noted that a fit of the
nucleon spin and flavor structure data in the two parameter broken-U(3)
chiral quark model favors the singlet to octet meson-quark coupling ratio to
be in the neighborhood of minus one. Thus we can interpret the result in (%
\ref{result}) as to have some phenomenological support. This in turn
buttresses the idea of a vanishing singlet axial charge of (\ref
{vanishing-gA}).

\section{Concluding remarks}

In this note we have provided further arguments to support the
target-independent suppression
  of the singlet axial vector charge. A
perturbative version of this result reflects the reasonable result
of heavy quark decoupling. In the pole dominance approximation, it leads to
a singlet meson-quark coupling with the opposite sign of the octet coupling,
a result 
which coincides with that of the chiral quark model phenomenology.

One of us (N.I.K.) is sincerely thankful to Prof.A. Di Giacomo
for warm hospitality at Universit\`a di Pisa and INFN for support.

\end{document}